\documentclass[12pt,a4paper]{article}
\usepackage{amssymb,amsmath}
\usepackage{epsf}
\usepackage[dvips]{epsfig}
\usepackage{lscape,multicol}

\def \un {\mbox{{\rm{\textbf{{1}}}}}}
\def \zero {\mbox{{\rm{\textbf{{0}}}}}}
\newcommand{\axiome}[1] {{\rm{\textbf{{#1}}}}}
\newcommand{\refe}[1] {(\ref{#1})}

\newcommand{\Sipos} {$\mbox{\v{S}ipo\v{s}}$}

\def\reel { {\rm  I\!R} }
\def\comp {{\mathchoice {\setbox0=\hbox{$\displaystyle\rm C$}\hbox{\hbox
to0pt{\kern0.4\wd0\vrule height0.96\ht0\hss}\box0}}
{\setbox0=\hbox{$\textstyle\rm C$}\hbox{\hbox
to0pt{\kern0.4\wd0\vrule height0.96\ht0\hss}\box0}}
{\setbox0=\hbox{$\scriptstyle\rm C$}\hbox{\hbox
to0pt{\kern0.4\wd0\vrule height0.96\ht0\hss}\box0}}
{\setbox0=\hbox{$\scriptscriptstyle\rm C$}\hbox{\hbox
to0pt{\kern0.4\wd0\vrule height0.96\ht0\hss}\box0}}}}
\def\zent {{\mathchoice {\hbox{$\sf\textstyle Z\kern-0.4em Z$}}
{\hbox{$\sf\textstyle Z\kern-0.4em Z$}}
{\hbox{$\sf\scriptstyle Z\kern-0.3em Z$}}
{\hbox{$\sf\scriptscriptstyle Z\kern-0.2em Z$}}}}

\newcommand{\pp}[1] { \left( {#1} \right)  }
\newcommand{\pa}[1] { \left\{ {#1} \right\}  }
\newcommand{\pr}[1] { \left[ {#1} \right]  }

\def\dep {\hspace*{-\parindent}}
\def\proof { \dep {\bf Proof :} }
\def\endproof {\rule{2mm}{2mm} \vspace*{3mm}}
\newtheorem{theorem}{Theorem}
\newtheorem{lemma}{Lemma}
\newtheorem{defin}{Definition}

\title {The Choquet integral for the aggregation of interval scales in 
	multicriteria decision making\footnotemark[1]
        \footnotetext[1]{This
        paper is an extended and revised version of a paper presented in the EUROFUSE 2001 conference {\protect\cite{Blab}}.}
 	\footnotetext[2]{On leave from Thales Research \& Technology, domaine de Corbeville,
	91404 Orsay cedex, France} }

\author{Christophe Labreuche  \\ Thales Research \& Technology \\
     Domaine de Corbeville  \\ 91404 Orsay cedex, France
 	\and  Michel Grabisch \footnotemark[2]
    \\ University of Paris VI \\ 4 place Jussieu \\ 75252 Paris, France}

\begin{document}

\maketitle

\begin{abstract}
This paper addresses the question of which models fit with information concerning
the preferences of the decision maker over each attribute, and his preferences
about aggregation of criteria (interacting criteria).
We show that the conditions induced by these information plus some intuitive
conditions lead to a unique possible aggregation operator: the Choquet integral.

\end{abstract}

\section{Introduction}

Let us consider a decision problem that depends on $n$ points of views described by the attributes
$X_1$, $\ldots$, $X_n$. The attributes can be any set. They can be of 
cardinal nature (for instance, the maximum speed of a car) or of ordinal nature 
(for instance, the color \{red,blue,$\ldots$\}).
We wish to model the preferences $\succeq$ of the decision maker (DM)
over acts, that is to say over elements of $X=X_1\times\cdots\times X_n$. 
For any $x\in X$, we use the following notation: $x=(x_1,\ldots,x_n)$
where $x_1\in X_1$, $\ldots$, $x_n\in X_n$.
A classical way is to model $\succeq$ with the help of an overall utility function
$u:X\rightarrow\reel$ \cite{Bkran}:
\begin{equation}
  \forall x,y\in X \ \ , \ \ x\succeq y \ \Leftrightarrow \ u\pp{x} \geq u\pp{y} \ .
\label{E1}
\end{equation}
The overall utility function $u$ is often taken in the following way \cite{Bkran}
\[ u\pp{x}=F\pp{u_1(x_1),\ldots,u_n(x_n)} \ ,
\]
where the $u_i$'s are the so called {\em utility functions}, and $F$ is 
the aggregation function. The main question is then how to determine the
$u_i$'s and $F$. The utility functions and the aggregation function
are generally constructed in two separate steps. Each criterion is
considered separately in the first step. The utility function $u_i$ represents
the preferences of the DM over the attribute $X_i$. The Macbeth approach
\cite{Bmac,Bmac1} provides a methodology to construct the $u_i$'s as 
scales of difference (see section \ref{S5a}). In the second 
step, all criteria are considered together. As example of aggregation functions, 
one can find the weighted sum, the Choquet integral or the Sugeno integral.
When $F$ is a Choquet integral, several methods for the determination of the
fuzzy measure are available. For instance, linear methods \cite{Bma2},
quadratic methods \cite{Bg4,Bg3} and heuristic-based methods \cite{Bg2}
are available in the literature.
For the Sugeno integral, heuristics \cite{Bg5} and methods based 
on fuzzy relations \cite{Brico,Brico1} can be found.
Unfortunately, these methods address only the problem of constructing the
aggregation function and refer more to learning procedures than to
true decision making approaches. The way the first step is dealt with is generally not
explained, especially with elaborate aggregators such as fuzzy integrals.

Yet, the two steps are intimately related. For instance, the notion of utility
function has no {\em absolute} meaning. One cannot construct 
the $u_i$'s without any a priori knowledge of what kind of $F$ will be considered.
However, when constructing the $u_i$'s, $F$ is not already known. As
a consequence, the construction of the utility functions is 
more complicated than what it seems.
One method addresses both points in a way that is satisfactory in the
measurement theory standpoint: this is the Macbeth approach. However, it is
restricted to the weighted sum. In such model, there is no interaction
between criteria. The aim of this paper is to extend the Macbeth approach
in such a way that interaction between criteria is allowed.

\bigskip

The $u_i$'s and $F$ shall be determined from some information obtained from the DM about
his preference over acts, e.g. elements of $X$. The information we will
consider in this paper can be seen as a generalization of the information needed in
the Macbeth technology \cite{Bmac}. As in Macbeth, the information is based on the
introduction of two absolute reference levels over each attribute, and 
the determination of scales of difference, which, put together, ensures commensurateness. 
We will consider more general information than in the Macbeth approach in such
a way that interaction between criteria will be allowed. 
We address then the question of which models fit with this information. The
information we have and the measurement conditions imply some 
conditions on the aggregation function $F$. As an example, 
one condition is that $F$ shall enable the construction of the $u_i$'s. Some conditions
given on specific acts are naturally extended to wider sets of acts.
These conditions rule out a wide range of families of aggregators. We see finally
which aggregation functions comes up. 
With these new properties, we show that the only possible model is
the Choquet integral.
We obtain an axiomatic representation of the Choquet integral that is
a weak version of a result shown by J.L. Marichal \cite{Bma1}.

We already addressed the same kind of problem for ratio scales in \cite{Bg1}.
For ratio scales, the \Sipos{} integral seems to be the right aggregation
function. The other difference between this paper and \cite{Bg1} is that
we restricted ourself only to the necessary information to construct the
model (preferences over each attribute and the aggregation of criteria)
and do not add any a priori assumption as in \cite{Bg1}.

\section{Available information}

The set of all criteria is denoted by $N=\pa{1,\ldots,n}$.
Considering two acts $x,y\in X$ and $A\subset N$, we use the notation
$(x_A,y_{-A})$ to denote the act $z\in X$ such that $z_i=x_i$ if $i\in A$
and $z_i=y_i$ otherwise.

It is well-known in MCDM that the preference relation $\succeq$ can be modeled
by an overall utility function $u$ (see \refe{E1}) if and only if all attributes are
set commensurate in some way \cite{Bkran}. The notion of commensurateness hinges on
the idea that we shall not try to aggregate directly attributes but rather 
aggregate values that represent the same kind of quantity. This is
the commensurate scale. 

The modeling of the preference relation of the
DM depends entirely on what type of scale the DM can work on. Here, we assume
that the DM can handle a scale of difference. This means that, on top of being
able to rank two acts $x,y\in X$, the DM can also give an assessment of the 
difference $u(x)-u(y)$ between the overall utility of $x$ and $y$. 
In other words, if $x\succeq y$, the DM can give the intensity with which $x$ is preferred to $y$.
We assume furthermore that the underlying scale is a bounded unipolar scale.
It is bounded from above and below. In this case, the commensurate
scale depicts the satisfaction degree of the DM over attributes.
The satisfaction degree is typically a number belonging to the interval
$[0,1]$. The two bounds $0$ and $1$ have a special meaning. The $0$ satisfaction 
value is the value that is considered completely unacceptable by the DM.
The $1$ satisfaction value is the value that is considered perfectly 
satisfactory by the DM. We assume that these two values $0$ and $1$ of the satisfaction
scale can be identified by two particular elements $\zero_i$ et $\un_i$ for
each attribute $X_i$. These two particular elements have an absolute meaning
throughout the attributes. We assume that $\zero_i$ is the worst element of $X_i$, that is to say
\begin{equation}
  \forall x_i\in X_i \ \ , \ \pp{x_i,\zero_{-i}} \succeq \pp{\zero_i,\zero_{-i}} \ .
\label{E2}
\end{equation}
Similarly, $\un_i$ is the best element of $X_i$~:
\begin{equation}
  \forall x_i\in X_i \ \ , \ \pp{x_i,\zero_{-i}} \preceq \pp{\un_i,\zero_{-i}} \ .
\label{E2a}
\end{equation}
 The introduction of $\zero_i$ and $\un_i$ enables us to
construct intra-criterion and inter-criteria information.

\subsection{Intra-criterion information}
For each criterion $i$, the mapping (denoted by $u_i$) from the attribute $X_i$
to the satisfaction scale $[0,1]$ must be explicited. This corresponds to the
preferences of the DM over each attribute. Commensurateness implies that
the elements of one attribute shall be compared to the elements of any other
scale. Taking a simple example involving two criteria (for instance consumption 
and maximal speed), this amounts to know whether the DM prefers a consumption of
$5\:$liters/$100$km to a maximum speed of $200\:$km/h. This does not generally make 
sense to the DM, so that he is not generally able to make this comparison directly.
In order to solve this problem out, the Macbeth approach is based on the idea that 
a scale of difference is constructed separately on each attribute. A scale
of difference is given up to two degrees of freedom. Fixing two points on the
scale determines entirely the scale. These two points are chosen in order
to enforce the overall commensurateness. As a consequence, these two points 
are the only elements of the attribute that have to be compared to the elements
of the other attributes. These elements are actually the $\zero_i$'s and $\un_i$'s.
All the $\zero_i$'s have the same meaning~: $u_1(\zero_1)=\cdots=u_n(\zero_n)$.
Similarly, $u_1(\un_1)=\cdots=u_n(\un_n)$.

It is natural for a DM to give his preferences over acts. On the other hand,
the DM shall not be asked information directly on the parameters of the model.
Therefore, it is not assumed that the DM can isolate attributes and give
information regarding directly the $u_i$'s.
To this end, as in the Macbeth methodology, we consider 
the subset $X\rfloor_i$ (for $i\in N$) of $X$ defined by
\[ X\rfloor_i := \pa{ \pp{x_i,\zero_{-i}} \ , \ x_i\in X_i } \ .
\]
We ask the DM not only the ranking of the elements of $X\rfloor_i$ but also 
the difference of satisfaction degree between pairs of elements of $X\rfloor_i$.
From this, $u_i$ can be defined by~:
\begin{description}
\item[(Intra$_a$)] $\forall x_i,y_i\in X_i$, 
$u_i(x_i)\geq u_i(y_i) \Leftrightarrow \pp{x_i,\zero_{-i}}\succeq \pp{y_i,\zero_{-i}}$.
\item[(Intra$_b$)] $\forall x_i,y_i,w_i,z_i\in X_i$ such that $u_i(x_i)>u_i(y_i)$
and $u_i(w_i)>u_i(z_i)$, we have
\[ \frac{u_i(x_i)-u_i(y_i)}{u_i(w_i)-u_i(z_i)} = k(x_i,y_i,w_i,z_i) 
       \ , \ \ k(x_i,y_i,w_i,z_i)\in\reel^+
\]
if and only if the difference of satisfaction degree that the DM feels between
$\pp{x_i,\zero_{-i}}$ and $\pp{y_i,\zero_{-i}}$ is $k(x_i,y_i,w_i,z_i)$ times as big as the
difference of satisfaction  between $\pp{w_i,\zero_{-i}}$ and $\pp{z_i,\zero_{-i}}$.
\item[(Intra$_c$)] $u_i(\zero_i)=0$ and $u_i(\un_i)=1$.
\end{description}
Condition {\bf (Intra$_b$)} means that $k(x_i,y_i,w_i,z_i)=
\frac{u\pp{x_i,\zero_{-i}}-u\pp{y_i,\zero_{-i}}}{u\pp{w_i,\zero_{-i}}-u\pp{z_i,\zero_{-i}}}$.
The $u_i$ correspond to a scale of difference. Such a scale is always given up to
a shift and a dilation. Condition {\bf (Intra$_c$)} fixes these two degrees 
of freedom. 

\medskip

In order to be able to construct a unique scale from {\bf (Intra$_a$)}, {\bf (Intra$_b$)}
and {\bf (Intra$_c$)}, some consistency assumptions shall be made~:
\begin{description}
\item[(Intra$_d$)] $\forall x_i,y_i,w_i,z_i,r_i,s_i\in X_i$ such that $u_i(x_i)>u_i(y_i)$,
$u_i(w_i)>u_i(z_i)$ and $u_i(r_i)>u_i(s_i)$,
\[ k(x_i,y_i,w_i,z_i) \times k(w_i,z_i,r_i,s_i) = k(x_i,y_i,r_i,s_i) \ .
\]
\end{description}

\subsection{Inter-criteria information}
We are concerned here in the determination of 
the aggregation function $F$. When $F$ is taken as a weighted sum, it is
enough to obtain from the DM some information regarding the importance of each
criterion. This is in particular the case of the Macbeth methodology
\cite{Bmac}. In the Macbeth approach, the following set of acts
is introduced 
\[ {\tilde{X}}:= \pa{\pp{\zero_N}} \bigcup
  \pa{\pp{\un_i,\zero_{-i}}  \:,\  i\in N} \ ,
\]
where $\pp{\zero_N}$ denotes the alternatives that is unacceptable on every
criteria.
On top of giving his preferences on the elements of ${\tilde{X}}$, the DM provides
some information on the difference of satisfaction between any two elements
of ${\tilde{X}}$. Thanks to that information, an interval scale that represents
the satisfaction degree of the elements of ${\tilde{X}}$ can be constructed,
provided the information is consistent.
Then, the importance of criterion $i$ is defined as being proportional
to the difference of the satisfaction degrees between 
$\pp{\un_i,\zero_{-i}}$ and $\pp{\zero_N}$.
The constant of proportionality is fixed from the property that all the
importances shall sum up to one.

\medskip

In order to generalize the Macbeth approach,
we introduce the following set $X\rceil_{\pa{0,1}}$ of acts~:
\[ X\rceil_{\pa{0,1}} := \pa{ \pp{\un_A,\zero_{-A}} \ , \ A\subset N} \ .
\]
As before, we ask information from which one can obtain a satisfaction
scale defined on $X\rceil_{\pa{0,1}}$. Define ${\mathcal P}(N)$ as the
set of all subsets of $N$. Let $\mu:{\mathcal P}(N)\rightarrow [0,1]$
defined by
\begin{description}
\item[(Inter$_a$)] $\forall A,B\subset N$, 
$\mu(A)\geq \mu(B) \Leftrightarrow \pp{\un_A,\zero_{-A}}\succeq \pp{\un_B,\zero_{-B}}$.
\item[(Inter$_b$)] $\forall A,B,C,D\subset N$ such that $\mu(A)>\mu(B)$
and $\mu(C)>\mu(D)$, we have
\[ \frac{\mu(A)-\mu(B)}{\mu(C)-\mu(D)} = k(A,B,C,D) \ , \ \ k(A,B,C,D)\in\reel^+
\]
if and only if the difference of satisfaction degree that the DM feels between
$\pp{\un_A,\zero_{-A}}$ and $\pp{\un_B,\zero_{-B}}$ is $k(A,B,C,D)$ times as big as the
difference of satisfaction  between $\pp{\un_C,\zero_{-C}}$ and $\pp{\un_D,\zero_{-D}}$.
\item[(Inter$_c$)] $\mu(\emptyset)=0$ and $\mu(N)=1$.
\end{description}

The last condition is rather natural since $\mu(\emptyset)=0$ means that the act 
which is completely unacceptable on all attributes is also completely 
unacceptable as a whole, and $\mu(N)=1$ means that the act which is perfectly satisfactory
on all attributes is also perfectly satisfactory as a whole. This depicts the
idea of commensurateness. 

\medskip

In order to be able to construct a unique scale from {\bf (Inter$_a$)}, {\bf (Inter$_b$)}
and {\bf (Inter$_c$)}, some consistency assumptions shall be made~:
\begin{description}
\item[(Inter$_d$)] $\forall A,B,C,D,E,F\subset N$ such that $\mu(A)>\mu(B)$,
$\mu(C)>\mu(D)$ and $\mu(E)>\mu(F)$
\[ k(A,B,C,D) \times k(C,D,E,F) = k(A,B,E,F) \ .
\]
\end{description}

\subsection{Measurement conditions}

The $u_i$'s and $\mu$ correspond to scales of difference. Considering for
instance $u_i$, this means that the only type of quantity that
makes sense is ratios of the form $\frac{u_i(x_i)-u_i(y_i)}{u_i(w_i)-u_i(z_i)}$.
Henceforth, it would have been possible to replace condition {\bf (Intra$_c$)}
by  $u_i(\zero_i)=\beta$ and $u_i(\un_i)=\alpha+\beta$, with $\alpha>0$ and 
$\beta\in\reel$. In other words, we could replace $u_i$ given by conditions 
{\bf (Intra$_a$)}, {\bf (Intra$_b$)} and {\bf (Intra$_c$)} by $\alpha u_i+\beta$.
Since the $u_i$'s map the attributes onto a single commensurate scale, one must
change all $u_i$'s in $\alpha u_i+\beta$ with the same $\alpha$ and $\beta$ 
(one cannot change only one scale). The measurement conditions for scales of difference imply
that the preference relation $\succeq$ and the ratios $\frac{u(x)-u(y)}{u(z)-u(t)}$
for all $x,y,z,t\in X$ shall not be changed if all the $u_i$'s are changed into $\alpha u_i+\beta$ 
with $\alpha>0$ and $\beta\in\reel$. We assume in this paper that this property
holds only for the acts that we are considering in our construction,
that is to say for $x,y,z,t\in X\rfloor_i$ and $x,y,z,t\in X\rceil_{\pa{0,1}}$.
This leads to the following requirement~:
\begin{description}
\item[(Intra$_e$)] The preference relation $\succeq$ and the $\frac{u(x)-u(y)}{u(z)-u(t)}$
for $x,y,z,t\in X\rfloor_i$ and for $x,y,z,t\in X\rceil_{\pa{0,1}}$
shall not be changed if all the $u_i$'s are changed into $\alpha u_i+\beta$ 
with $\alpha>0$ and $\beta\in\reel$.
\end{description}
This condition is no more than the requirement that the $u_i$'s are commensurate
scales of differences.

\bigskip

Looking at $X\rceil_{\pa{0,1}}$ as a generalization of the set $\tilde{X}$ used in the 
Macbeth approach, $\mu\pp{\pa{i}}$ represents some kind of {\em importance 
of criterion $i$}. More precisely, the term $\mu\pp{\pa{i}}$ corresponds 
to the difference of the satisfaction degrees between 
$\pp{\un_i,\zero_{-i}}$ and $\pp{\zero_N}$.
Generalizing this to coalitions of any cardinality, $\mu(A)$ corresponds in
fact to the difference of the satisfaction degrees between the alternatives
$\pp{\un_A,\zero_{-A}}$ and $\pp{\zero_N}$.
Applying this to $A=\emptyset$, the value $\mu\pp{\emptyset}$ shall always
be equal to zero, whatever the interval scale attached to ${\tilde{X}}$
may be. If the interval scale attached to ${\tilde{X}}$ is
changed to another interval scale, the values of $\mu(A)$ could vary,
except $\mu\pp{\emptyset}$ that will always vanish. More precisely,
$\mu$ will be replaced by $\gamma \mu$, with $\gamma\in \reel$. Henceforth, $\mu$
corresponds to a ratio scale. As a consequence, it would be possible to
replace condition {\bf (Inter$_c$)} by $\mu(\emptyset)=0$ and $\mu(N)=\gamma$,
with $\gamma\in \reel$. As previously, the following requirement is imposed~:
\begin{description}
\item[(Inter$_e$)] The preference relation $\succeq$ and the $\frac{u(x)-u(y)}{u(z)-u(t)}$
for $x,y,z,t\in X\rfloor_i$ and for $x,y,z,t\in X\rceil_{\pa{0,1}}$
shall not be changed if $\mu$ is changed into $\gamma\mu$ with $\gamma\in\reel$.
\end{description}

\section{Conditions on the model}
In measurement theory, it is classical to split the overall evaluation model
$u$ into two parts \cite{Bkran}: the utility functions (that map the attributes onto a single
satisfaction degree scale), and the aggregation function (that aggregates
commensurate scales). 
The function $u_i$ corresponds to satisfaction degrees over criterion $i$.
Thanks to assumptions {\bf (Intra$_a$)} and  {\bf (Intra$_b$)}, $u_i$ are scales of difference.
These scales are commensurate by condition {\bf (Intra$_c$)}.
The term $\mu(A)$ represents the importance that the
DM gives to the coalition $A$ in the DM process. Consequently, it is natural
to write $u$ as follows:
\begin{equation}
 u\pp{x}=F_\mu\pp{u_1(x_1),\ldots,u_n(x_n)} \ ,
\label{E3}
\end{equation}
where $F_\mu$ is the aggregation operator. $F_\mu$ depends on $\mu$ in a 
way that is not known for the moment. From now on, we assume that the preferences 
of the DM can be modeled by $u$ given by \refe{E3}.

\bigskip

As example of aggregation operators, $F_\mu$ could be a weighted sum
\[  F_\mu\pp{u_1,\ldots,u_n}=\sum_{i=1}^n \alpha_i \: u_i \ .
\]
For a normalized weighted sum (i.e. when the weights $\alpha_i$ sum up to one),
one clearly has $F_\mu(\beta,\ldots,\beta) = \beta$ for any $\beta$.
This property is explained by the fact that $F_\mu$ aggregates commensurate scales.
It can be naturally generalized to any $F_\mu$. The condition
that the weighted sum is normalized becomes that $\mu$ is normalized, that
is to say $\mu(N)=1$. Henceforth, it is natural to assume that (whenever $\mu$ satisfies 
$\mu(N)=1$)
\[  F_\mu(\beta,\ldots,\beta) = \beta \ \ , \ \forall\beta\in [0,1] \ .
\]
Due to condition {\bf (Intra$_c$)} and relations \refe{E2} and \refe{E2a},
$u_i(x_i)\in [0,1]$ for any $x_i\in X_i$. This is why $\beta\in[0,1]$
in previous relation. However, condition {\bf (Intra$_e$)}
implies that $u_i(x_i)$ could take virtually any real value. Hence,
$F_\mu$ shall be defined on $\reel^n$. This shows that previous relation
on $F_\mu$ shall hold for any $\beta\in\reel$~:
\[  F_\mu(\beta,\ldots,\beta) = \beta \ \ , \ \forall\beta\in \reel \ .
\]
Consider now the case when $\mu(N)\not=1$. Since $\mu$ corresponds to
a ratio scale, it is reasonable to assume that $\mu$ acts as a dilation
of the overall evaluation scale of $u$. In other words, when $\mu(N)\not= 1$, 
we should have 
\begin{equation}
  F_\mu(\beta,\ldots,\beta) = \beta \mu(N) \ \ , \ \forall\beta\in\reel \ .
\label{E8}
\end{equation}

\subsection{Intra-criterion information}

\begin{lemma}
If $u$ satisfies \refe{E3}, and if conditions {\bf (Intra$_a$)}, {\bf (Intra$_b$)},
{\bf (Intra$_c$)}, {\bf (Intra$_d$)}, {\bf (Intra$_e$)}, {\bf (Inter$_a$)}, {\bf (Inter$_b$)},
{\bf (Inter$_c$)}, {\bf (Inter$_d$)} and {\bf (Inter$_e$)} are fulfilled, then
for all $a_i,b_i,c_i,d_i\in [0,1]$, and for all $\alpha>0$, $\gamma,\beta\in\reel$,
\begin{equation}
  \frac{F_{\gamma\mu}\pp{\alpha a_i+\beta,\beta_{-i}}-F_{\gamma\mu}\pp{\alpha b_i+\beta,\beta_{-i}}}
   {F_{\gamma\mu}\pp{\alpha c_i+\beta,\beta_{-i}}-F_{\gamma\mu}\pp{\alpha d_i+\beta,\beta_{-i}}}
  = \frac{a_i-b_i}{c_i-d_i} \ .
\label{E5}
\end{equation}
\label{L2}
\end{lemma}

\proof
$u\pp{x_i,\zero_{-i}}$ and $u_i(x_i)$ correspond to two possible
scales of difference related to the same act $\pp{x_i,\zero_{-i}}\in X\rfloor_i$.
Henceforth
\[ \frac{u\pp{x_i,\zero_{-i}}-u\pp{y_i,\zero_{-i}}}{u\pp{w_i,\zero_{-i}}-u\pp{z_i,\zero_{-i}}}
   =\frac{u_i(x_i)-u_i(y_i)}{u_i(w_i)-u_i(z_i)} \ ,
\]
which gives
\begin{equation}
  \frac{F_\mu\pp{u_i(x_i),u_{-i}(\zero_{-i})}-F_\mu\pp{u_i(y_i),u_{-i}(\zero_{-i})}}
       {F_\mu\pp{u_i(w_i),u_{-i}(\zero_{-i})}-F_\mu\pp{u_i(z_i),u_{-i}(\zero_{-i})}}
  = \frac{u_i(x_i)-u_i(y_i)}{u_i(w_i)-u_i(z_i)} \ .
\label{E4}
\end{equation}
From {\bf (Intra$_e$)}, one might change all $u_j$ in $\alpha u_j +\beta$ 
at the same time without any change in \refe{E4}.
In this case, the utility functions will take values in the interval
$[\beta,\alpha+\beta]$. From {\bf (Inter$_e$)}, one might change $\mu$ into
$\gamma \mu$ ($\gamma\in\reel$) without any change in \refe{E4}. Consequently, one
must have
\begin{eqnarray*}
\lefteqn{ \frac{F_{\gamma \mu}\pp{\alpha u_i(x_i)+\beta,\alpha u_{-i}(\zero_{-i})+\beta}
-F_{\gamma \mu}\pp{\alpha u_i(y_i)+\beta,\alpha u_{-i}(\zero_{-i})+\beta}}
       {F_{\gamma \mu}\pp{\alpha u_i(w_i)+\beta,\alpha u_{-i}(\zero_{-i})+\beta}
-F_{\gamma \mu}\pp{\alpha u_i(z_i)+\beta,\alpha u_{-i}(\zero_{-i})+\beta}}  }  \\
 & & \ \ \ \ \ \ \ \ \ \ \ \ \ \ \ \ \ \ \ \ \ \ \ \ \ \ \ \ \ \ \ \ \ \ \ 
  \ \ \ \ \ \ \ \ \ \ \ \ \ 
  = \frac{\pp{\alpha u_i(x_i)+\beta}-\pp{\alpha u_i(y_i)+\beta}}
         {\pp{\alpha u_i(w_i)+\beta}-\pp{\alpha u_i(z_i)+\beta}} \ .
\end{eqnarray*}
Now, setting $a_i=u_i(x_i)$, $b_i=u_i(y_i)$, $c_i=u_i(w_i)$ and $d_i=u_i(z_i)$,
we have $a_i,b_i,c_i,d_i \in [0,1]$ by \refe{E2}, \refe{E2a} and {\bf (Intra$_c$)}.
Hence, the lemma is proved.
\endproof

\subsection{Inter-criteria information}

\begin{lemma}
If $u$ satisfies \refe{E3}, and if conditions {\bf (Intra$_a$)}, {\bf (Intra$_b$)},
{\bf (Intra$_c$)}, {\bf (Intra$_d$)}, {\bf (Intra$_e$)}, {\bf (Inter$_a$)}, {\bf (Inter$_b$)},
{\bf (Inter$_c$)}, {\bf (Inter$_d$)} and {\bf (Inter$_e$)} are fulfilled, then
for any $\alpha>0$ and $\gamma,\beta\in\reel$, it holds that
\begin{equation}
\frac{F_{\gamma\mu}\pp{(\alpha+\beta)_A,\beta_{-A}}-F_{\gamma\mu}\pp{(\alpha+\beta)_B,\beta_{-B}}}
{F_{\gamma\mu}\pp{(\alpha+\beta)_C,\beta_{-C}}-F_{\gamma\mu}\pp{(\alpha+\beta)_D,\beta_{-D}}}
  = \frac{\mu(A)-\mu(B)}{\mu(C)-\mu(D)} \ .
\label{E7}
\end{equation}
\label{L3}
\end{lemma}

\proof
By conditions \axiome{(Inter$_a$)} and \axiome{(Inter$_b$)},
$u\pp{\un_A,\zero_{-A}}$ and $\mu(A)$ correspond to two possible interval
scales related to the act $\pp{\un_A,\zero_{-A}}\in X\rceil_{\pa{0,1}}$.
Hence
\[ \frac{u\pp{\un_A,\zero_{-A}}-u\pp{\un_B,\zero_{-B}}}
     {u\pp{\un_C,\zero_{-C}}-u\pp{\un_D,\zero_{-D}}}
  =\frac{\mu(A)-\mu(B)}{\mu(C)-\mu(D)} \ ,
\]
which gives
\begin{equation}
  \frac{F_{\mu}\pp{u_A(\un_A),u_{-A}(\zero_{-A})}-F_{\mu}\pp{u_B(\un_B),u_{-B}(\zero_{-B})}}
  {F_{\mu}\pp{u_C(\un_C),u_{-C}(\zero_{-C})}-F_{\mu}\pp{u_D(\un_D),u_{-D}(\zero_{-D})}}
  = \frac{\mu(A)-\mu(B)}{\mu(C)-\mu(D)} \ .
\label{E6}
\end{equation}
As previously, from {\bf (Intra$_e$)} and {\bf (Inter$_e$)}, 
one can change $u_j$ in $\alpha u_j +\beta$, and $\mu$ in
$\gamma \mu$ without any change in \refe{E6}~:
\begin{eqnarray*}
\lefteqn{  \frac{F_{\gamma\mu}\pp{\alpha u_A(\un_A)+\beta,\alpha u_{-A}(\zero_{-A})+\beta}
-F_{\gamma\mu}\pp{\alpha u_B(\un_B)+\beta,\alpha u_{-B}(\zero_{-B})+\beta}}
  {F_{\gamma\mu}\pp{\alpha u_C(\un_C)+\beta,\alpha u_{-C}(\zero_{-C})+\beta}
-F_{\gamma\mu}\pp{\alpha u_D(\un_D)+\beta,\alpha u_{-D}(\zero_{-D})+\beta}}  } \\
 & & \ \ \ \ \ \ \ \ \ \ \ \ \ \ \ \ \ \ \ \ \ \ \ \ \ \ \ \ \ \ \ \ \ \ \ 
  \ \ \ \ \ \ \ \ \ \ \ \ \ \ \ \ \ \ \ \ \ \ \ \ \ \ \ \ \ \ 
  = \frac{\gamma\mu(A)-\gamma\mu(B)}{\gamma\mu(C)-\gamma\mu(D)} \ .
\end{eqnarray*}
Hence the lemma is proved.
\endproof

\section{Generalized conditions on the model}

\begin{lemma}
If $u$ satisfies \refe{E3}, and if conditions \refe{E8}, {\bf (Intra$_a$)}, {\bf (Intra$_b$)},
{\bf (Intra$_c$)}, {\bf (Intra$_d$)}, {\bf (Inter$_a$)}, {\bf (Inter$_b$)},
{\bf (Inter$_c$)} and {\bf (Inter$_d$)} are fulfilled, then
we have for any $\eta, \beta\in\reel$
\begin{equation}
  F_{\gamma\mu+\delta\mu'}\pp{\eta_A,\beta_{-A}} 
  = \gamma F_\mu\pp{\eta_A,\beta_{-A}}
  +\delta F_{\mu'}\pp{\eta_A,\beta_{-A}} \ .
\label{E10}
\end{equation}
\label{L4}
\end{lemma}

\proof
Taking \refe{E7} with $\gamma=1$, $B=D=\emptyset$ and $C=N$, it holds that
\begin{equation}
  F_\mu\pp{ (\alpha+\beta)_A, \beta_{-A} } = \alpha \mu(A) + \beta \mu(N) \ .
\label{E9}
\end{equation}
This relation holds for any $\mu$ satisfying $\mu(\emptyset)=0$.
Consequently, replacing $\mu$ by $\gamma \mu + \delta \mu'$, we obtain
for any $\beta\in\reel$, $\alpha>0$
\[  F_{\gamma\mu+\delta\mu'}\pp{(\alpha+\beta)_A,\beta_{-A}} 
  = \gamma F_\mu\pp{(\alpha+\beta)_A,\beta_{-A}}
  +\delta F_{\mu'}\pp{(\alpha+\beta)_A,\beta_{-A}} \ .
\]
Hence, setting $\eta=\alpha+\beta$, we obtain for any $\beta\in\reel$ and $\eta>\beta$,
\[  F_{\gamma\mu+\delta\mu'}\pp{\eta_A,\beta_{-A}} 
  = \gamma F_\mu\pp{\eta_A,\beta_{-A}}
  +\delta F_{\mu'}\pp{\eta_A,\beta_{-A}} \ .
\]
Replacing $A$ by $N \setminus A$ in previous relation, we obtain for 
any $\beta\in\reel$ and $\eta>\beta$,
\[  F_{\gamma\mu+\delta\mu'}\pp{\beta_A,\eta_{-A}} 
  = \gamma F_\mu\pp{\beta_A,\eta_{-A}}
  +\delta F_{\mu'}\pp{\beta_A,\eta_{-A}} \ .
\]
Putting together last two results, we get for any $\eta,\beta\in\reel$ 
with $\beta\not= \eta$
\[  F_{\gamma\mu+\delta\mu'}\pp{\eta_A,\beta_{-A}} 
  = \gamma F_\mu\pp{\eta_A,\beta_{-A}}
  +\delta F_{\mu'}\pp{\eta_A,\beta_{-A}} \ .
\]
Previous relation holds also when $\beta=\eta$ by \refe{E8}.
Hence, we have for any $\eta, \beta\in\reel$
\[  F_{\gamma\mu+\delta\mu'}\pp{\eta_A,\beta_{-A}} 
  = \gamma F_\mu\pp{\eta_A,\beta_{-A}}
  +\delta F_{\mu'}\pp{\eta_A,\beta_{-A}} \ .
\]
\endproof

\bigskip

We wish to generalize \refe{E10} in the following way
\[  F_{\gamma\mu+\delta\mu'}(x) = \gamma F_\mu(x) + \delta F_{\mu'}(x)
  \ \ , \ \forall x\in\reel^n, \ \forall \gamma,\delta\in\reel \ .
\]
We should have $F_{\gamma \mu}(x) = \gamma F_{\mu}(x)$ ($\gamma\in\reel$)
since $\mu$ corresponds to a ratio scale. One should have 
$F_{\frac{\mu+\mu'}{2}}(x)=\frac{1}{2}\pp{F_\mu(x)+F_{\mu'}(x)}$ for the
following reason. If two decision makers give the importances of the coalitions
($\mu$ and $\mu'$ respectively), the consensus of these two decision makers
can be performed by taking the mean value of these information (that is to say
$\frac{\mu+\mu'}{2}$). Then it is reasonable that the overall aggregation functions
equals the mean value of the aggregation for the two DM. Based on these 
considerations, we propose the following axiom~:

\begin{quote}
\axiome{Linearity wrt the Measure (LM)}: For all $x\in\reel^n$ and $\gamma,\delta\in\reel$,
\begin{equation}
 F_{\gamma\mu+\delta\mu'}(x) = \gamma F_\mu(x) + \delta F_{\mu'}(x) \ .
\label{E11}
\end{equation}
\end{quote}

This axiom cannot be completely deduced from the construction of the $u_i$'s and $\mu$. However,
Lemma \ref{L4} proves that formula \refe{E11} is obtained from the construction of
the $u_i$'s and $\mu$ for particular alternatives $x$.

\bigskip

Since $F_\mu$ aggregates satisfaction scales, it is natural to assume 
that $x\mapsto F_\mu(x)$ is increasing.
\begin{quote}
\axiome{Increasingness (In)}: $\forall x,x'\in\reel^n$,
\[ x_i\leq {x'}_i \: \forall i\in N \ \Rightarrow F_\mu(x)\leq F_\mu(x')
\]
\end{quote}

This axiom is not deduced from the construction of the $u_i$'s and $\mu$. It is a 
necessary requirement for $F_\mu$ to be an aggregation function
\cite{Bforou}.

\bigskip

Let us give now the last two axioms.
\begin{quote}
\axiome{Properly Weighted (PW)}: If $\mu$ satisfies condition \axiome{(Inter$_c$)},
then $F_\mu\pp{1_A,0_{-A}} = \mu(A)$, $\forall A\subset N$.
\end{quote}

\bigskip

\begin{quote}
\axiome{Stability for the admissible Positive Linear transformations (weak SPL)}: 
If $\mu$ satisfies condition \axiome{(Inter$_c$)}, then for all $A\subset N$, $\alpha>0$, 
and $\beta\in \reel$,
\[ F_\mu\pp{(\alpha+\beta)_A,\beta_{-A}}=\alpha F_\mu\pp{1_A,0_{-A}}+\beta
\]
\end{quote}
This axiom is called \axiome{(weak SPL)} since it is a weak version of
the axiom \axiome{(SPL)} introduced by J.L. Marichal \cite{Bma0,Bma1}~:
\begin{quote}
\axiome{(SPL)}: For all $x\in\reel^n$, $\alpha>0$, and $\beta\in \reel$,
\[ F_\mu\pp{\alpha x+\beta}=\alpha F_\mu\pp{x}+\beta
\]
\end{quote}

\begin{lemma}
Axioms \axiome{(PW)} and \axiome{(weak SPL)} can be deduced from the construction
of the $u_i$'s and $\mu$.
\label{L5}
\end{lemma}

\proof
Axiom \axiome{(PW)} is obtained from \refe{E9} with $\beta=0$ and $\alpha=1$.
Finally, using \refe{E9} and {\bf (PW)}, \axiome{(weak SPL)} holds.
\endproof

\bigskip

Next lemma shows that axioms \axiome{(LM)}, \axiome{(PW)} and \axiome{(weak SPL)}
contain the two relations \refe{E5} and \refe{E7}.
\begin{lemma}
Conditions \axiome{(LM)}, \axiome{(PW)} and \axiome{(weak SPL)}
imply relations \refe{E5} and \refe{E7}.
\label{L0}
\end{lemma}

\proof
Consider $F_\mu$ satisfying \axiome{(LM)}, \axiome{(PW)} and \axiome{(weak SPL)}. 
By \axiome{(LM)} and \axiome{(weak SPL)}, we have for any $\alpha>0$
and $a_i\in\reel^+$,
\[ F_{\gamma\mu}\pp{\alpha a_i+\beta,\beta_{-i}} 
  = \gamma F_{\mu}\pp{\alpha a_i+\beta,\beta_{-i}} 
  = \gamma\alpha a_i F_\mu\pp{1_i,0_{-i}} + \gamma\beta \ .
\]
Consequently, \refe{E5} is fulfilled. Condition \refe{E7} is clearly satisfied
from \axiome{(LM)}, \axiome{(PW)} and \axiome{(weak SPL)}.
\endproof

To end up this section, let us recall that the construction of the $u_i$'s (conditions
{\bf (Intra$_a$)}, {\bf (Intra$_b$)}, {\bf (Intra$_c$)}, {\bf (Intra$_d$)}), {\bf (Intra$_e$)})
and $\mu$ (conditions {\bf (Inter$_a$)}, {\bf (Inter$_b$)}, {\bf (Inter$_c$)},
{\bf (Inter$_d$)}) and {\bf (Inter$_e$)}, plus assumptions \refe{E3}
and \refe{E8} lead to axioms \axiome{(PW)} and \axiome{(weak SPL)}, and
partially to axiom \axiome{(LM)} (see Lemma \ref{L4}). Moreover, we saw that
axiom \axiome{(In)} is a very natural requirement.

\section{Expression of the aggregation function}

In this section, we wish to find which aggregation functions $F_\mu$ are characterized
by conditions \axiome{(LM)}, \axiome{(In)}, \axiome{(PW)} and \axiome{(weak SPL)}.
It can be noticed that the Choquet integral satisfies these conditions. Our
main result shows that this is the only aggregator satisfying these conditions.

\subsection{Background on the Choquet integral}

Thanks to condition \axiome{(Inter$_c$)} and axiom \axiome{(In)}, the
set function $\mu$ satisfies
\begin{enumerate}
\item[(i)] $\mu(\emptyset)=0$, $\mu(N)=1$.
\item[(ii)] $A\subset B\subset N$ implies $\mu(A)\leq \mu(B)$.
\end{enumerate}
These two properties define the so-called {\em fuzzy measures}.
As a consequence, $F_\mu$ could be a Choquet integral.

\bigskip

The Choquet integral \cite{Bforou,Bg6} is a generalization of the commonly used
weighted sum. Whereas the weighted sum is the discrete form of
the Lebesgue integral with an additive measure, we consider
as an aggregator the discrete form of the generalization
of Lebesgue's integral to the case of non-additive measures.
This generalization is precisely the Choquet integral.

\begin{defin}
Let $\mu$ be a fuzzy measure on $N$. The 
{\em discrete Choquet integral} of an element $x\in\reel^n$
with respect to $\mu$ is defined by
\begin{eqnarray*}
C_{\mu}(x) & = & \sum_{i=1}^n x_{\tau(i)} [ \mu\pp{\pa{\tau(i),\ldots,\tau(n)}} \\
  & & \ \ \ \ \ \ \ \ \ \ \ \    - \mu\pp{\pa{\tau(i+1),\ldots,\tau(n)}} ]
\end{eqnarray*}
where $\tau$ is a permutation satisfying 
$x_{\tau(1)} \leq \cdots \leq x_{\tau(n)}$.
\end{defin}

Besides, the Choquet integral can model typical human
behavior such as the veto. This operator is also able to
model the importance of criteria and the interaction between 
criteria. Conversely, the Choquet integral can be 
interpreted in term of the importance of criteria, the 
interaction between criteria, and veto \cite{Bg3,Bg6,Bforou}.

\subsection{Expression of $F_\mu$}
The result shown here is close to a result proved in \cite{Bma1}.
Before giving the final result, let us give an intermediate lemma.

\begin{lemma}
  If $F:\reel^n\rightarrow\reel$ satisfies \axiome{(In)} and \axiome{(weak SPL)},
$F(0_N)=0$, $F(1_N)=1$ and that $F(1_A,0_{-A})\in\pa{0,1}$ 
for all $A\subset N$, then $F\equiv C_\mu$, where
the fuzzy measure $\mu$ is given by $\mu(A)=F(1_A,0_{-A})$ for all $A\subset N$.
\label{L1}
\end{lemma}

\proof
Let $\sigma$ be a permutation of $N$. For $i\in N$, let 
\[
  \theta_\sigma(i):=F\pp{1_{\pa{\sigma(i),\ldots,\sigma(n)}},0_{\pa{\sigma(1),\ldots,\sigma(i-1)}}} \ .
\]
We have $\theta_\sigma(i)\in\pa{0,1}$. From \axiome{(In)}, there exists $k_\sigma\in N$ such that
\[ \theta_\sigma(i)=0 \mbox{ for } i\in\pa{k_\sigma+1,\ldots,n} \ \mbox{ and } \ 
   \theta_\sigma(i)=1 \mbox{ for } i\in\pa{1,\ldots,k_\sigma} \ .
\]
We have $k_\sigma\in\pa{1,\ldots,n-1}$ since $F(0_N)=0$ and $F(1_N)=1$.
For $x\in \reel^n$, the following notation $\pr{x_1,\ldots,x_n}_{\sigma^{-1}}$ denotes the
element $y\in \reel^n$ such that $y_i=x_{\sigma^{-1}(i)}$. With this notation, $x\in \reel^n$ reads
$x=\pr{x_{\sigma(1)},\ldots,x_{\sigma(n)}}_{\sigma^{-1}}$.

Let ${\mathcal{B}}_\sigma:=\pa{x\in\reel^n \ , \ x_{\sigma(1)}\leq \ldots \leq x_{\sigma(n)}}$ and
$x\in {\mathcal{B}}_\sigma$. From \axiome{(In)} and \axiome{(weak SPL)}, and since
$x_{\sigma(k_\sigma)}\leq x_{\sigma(n)}$, we have
\begin{eqnarray*}
 F(x) & = & F\pp{\pr{x_{\sigma(1)},\ldots,x_{\sigma(n)}}_{\sigma^{-1}}} \\
   & \leq & F([ \underbrace{x_{\sigma(k_\sigma)},\ldots,x_{\sigma(k_\sigma)}}_{k_\sigma\ {\rm times}},
       \underbrace{x_{\sigma(n)},\ldots,x_{\sigma(n)}}_{(n-k_\sigma)\ {\rm times}} ]_{\sigma^{-1}} ) \\
  & =& x_{\sigma(k_\sigma)} + \pp{x_{\sigma(n)}-x_{\sigma(k_\sigma)}}
    F([ \underbrace{0,\ldots,0}_{k_\sigma\ {\rm times}},
       \underbrace{1,\ldots,1}_{(n-k_\sigma)\ {\rm times}} ]_{\sigma^{-1}} ) \\
  & = & x_{\sigma(k_\sigma)} + \pp{x_{\sigma(n)}-x_{\sigma(k_\sigma)}}
     F\pp{1_{\pa{\sigma(k_\sigma+1),\ldots,\sigma(n)}},0_{\pa{\sigma(1),\ldots,\sigma(k_\sigma)}}} \\
  & = &  x_{\sigma(k_\sigma)} + \pp{x_{\sigma(n)}-x_{\sigma(k_\sigma)}} \theta_\sigma(k_\sigma+1)
    = x_{\sigma(k_\sigma)}
\end{eqnarray*}
since $\theta_\sigma(k_\sigma+1)=0$.
On the other hand, by \axiome{(In)},
\begin{eqnarray*}
 F(x) & = & F\pp{\pr{x_{\sigma(1)},\ldots,x_{\sigma(n)}}_{\sigma^{-1}}} \\
  & \geq & F([ \underbrace{x_{\sigma(1)},\ldots,x_{\sigma(1)}}_{(k_\sigma-1)\ {\rm times}},
     \underbrace{x_{\sigma(k_\sigma)},\ldots,x_{\sigma(k_\sigma)}}_{(n-k_\sigma+1)\ {\rm times}} ]_{\sigma^{-1}} ) \\
  & = & x_{\sigma(1)} + \pp{x_{\sigma(k_\sigma)}-x_{\sigma(1)}} 
     F\pp{1_{\pa{\sigma(k_\sigma),\ldots,\sigma(n)}},0_{\pa{\sigma(1),\ldots,\sigma(k_\sigma-1)}}} \\
  & = & x_{\sigma(1)} + \pp{x_{\sigma(k_\sigma)}-x_{\sigma(1)}} \theta_\sigma(k_\sigma)
  =  x_{\sigma(k_\sigma)}
\end{eqnarray*}
since $\theta_\sigma(k_\sigma)=1$. Hence we have
\[ F(x) = x_{\sigma(k_\sigma)} \ .
\]
Let us note that for the fuzzy measure $\mu$ defined by $\mu(S)=F(1_S,0_{-S})$, we have
\[
 C_\mu(x)=\sum_{i=1}^n x_{\sigma(i)} \pr{\mu(\pa{\sigma(i),\ldots,\sigma(n)})-\mu(\pa{\sigma(i+1),\ldots,\sigma(n)})} \ .
\]

Moreover, 
\begin{eqnarray*}
 \mu(\pa{\sigma(i),\ldots,\sigma(n)}) & = &
  F\pp{1_{\pa{\sigma(i),\ldots,\sigma(n)}},0_{\pa{\sigma(1),\ldots,\sigma(i-1)}}} \\
  & = &  \theta_\sigma(i)  =  \left\{ \begin{array}{l}
    1 \mbox{ if } i\in\pa{1,\ldots,k_\sigma} \\ 0 \mbox{ if } i\in\pa{k_\sigma+1,\ldots,n}
  \end{array} \right.
\end{eqnarray*}
Hence
\[ C_\mu(x)=x_{\sigma(k_\sigma)}=F(x) \ .
\]
\endproof

In Lemma \ref{L1}, \axiome{(In)} and \axiome{(weak SPL)} are sufficient to
obtain that $F$ is a Choquet integral only because the range of $F(1_A,0_{-A})$
belongs to $\pa{0,1}$. This is no more true in the general case.

\begin{theorem}
$F_\mu$ satisfies \axiome{(LM)}, \axiome{(In)}, \axiome{(PW)} and \axiome{(weak SPL)}
if and only if $F_\mu\equiv C_\mu$ in $\reel^n$.
\label{T1}
\end{theorem}

\proof
Clearly, the Choquet integral satisfies \axiome{(LM)}, \axiome{(In)}, \axiome{(PW)} 
and \axiome{(weak SPL)}.

Consider now $F_\mu$ satisfying \axiome{(LM)}, \axiome{(In)}, \axiome{(PW)} 
and \axiome{(weak SPL)}.
We write $\mu(A)=\sum_{B\subset N} m(B) u_B(A)$, where $m$ is the M\"{o}bius transform
$m(B)=\sum_{C\subset B} (-1)^{|B|-|C|} \mu(C)$, and $u_B$ is the unanimity
game, that is to say $u_B(A)$ equals $1$ if $B \subset A$ and $0$ otherwise.
Since $m(\emptyset)=0$, we have
$\mu(A)=\sum_{B\subset N\:,\: B\not=\emptyset} m(B) u_B(A)$
Thanks to \axiome{(LM)},
\[ F_\mu(x) = \sum_{B\subset N\:,\: B\not=\emptyset} m(B) F_{u_B}(x) \ ,
\]
where $u_B$ is a $\pa{0,1}$ valued fuzzy measure. By lemma \ref{L1}, for $B\not=\emptyset$,
there exists a fuzzy measure $v$ such that $F_{u_B}(x)=C_v(x)$ for all 
$x\in\reel^n$. Clearly, by \axiome{(PW)}, we have for $B\not=\emptyset$
\[ v(A)=F_{u_B}(1_A,0_{-A}) = u_B(A) \ .
\]
Hence 
\[ F_{u_B}(x)=C_{u_B}(x) \ \forall x\in\reel^n \ .
\]
By linearity of the Choquet integral with respect to the fuzzy measure, 
\[ \forall x\in\reel^n, F_\mu(x)=\sum_{B\subset N\:,\: B\not=\emptyset} m(B) C_{u_B}(x) =
  C_{\sum_{B\subset N\:,\: B\not=\emptyset} m(B) u_B}(x) = C_\mu(x) \ .
\]
\endproof

This theorem is a generalization of a result of J.L. Marichal \cite{Bma1}.
J.L. Marichal proved in \cite{Bma1} that the Choquet integral is the only 
aggregation function satisfying \axiome{(LM)}, \axiome{(In)}, \axiome{(PW)} 
and \axiome{(SPL)}.

\bigskip

Let us sum up the result we have shown.
From the construction of the $u_i$'s (conditions
{\bf (Intra$_a$)}, {\bf (Intra$_b$)}, {\bf (Intra$_c$)}, {\bf (Intra$_d$)}),
{\bf (Intra$_e$)}) and $\mu$ (conditions {\bf (Inter$_a$)}, {\bf (Inter$_b$)},
{\bf (Inter$_c$)}, {\bf (Inter$_d$)}), {\bf (Inter$_e$)}), and adding the two assumptions \refe{E3}
and \refe{E8}, we obtained axioms \axiome{(PW)} and \axiome{(weak SPL)}
(see Lemma \ref{L0}), and axiom \axiome{(LM)} was partially obtained 
(see Lemma \ref{L4}). Moreover, we saw that
axiom \axiome{(In)} is a very natural requirement.
Theorem \ref{T1} shows that there is only one model that fits with previous
information. This is the Choquet integral. Therefore, the Choquet integral
is the only model that is suitable with our construction.

\section{Discussion on the practical construction of the $u_i$'s and $\mu$}
\label{S5}

Some practical issues concerning the construction of $u_i$ and $\mu$ are considered
in this section.

\subsection{Intra-criterion information}
\label{S5a}

Let us investigate here how to construct in practice the utility function $u_i$.
We follow the Macbeth methodology \cite{Bmac,Bmac1}. The goal of this section
is not to give very precise details concerning the Macbeth approach. We refer to
references \cite{Bmac,Bmac1} for a more detailed explanation of the Macbeth approach.

For the sake of simplicity, assume that attribute $X_i$ has a finite set of values~:
$X_i=\pa{a^i_1,\ldots,a^i_{p_i}}$ with $a^i_1=\zero_i$ et $a^i_{p_i}=\un_i$. 
We aim at determining $u_i(a^i_j)$ for all $j\in\pa{1,\ldots,p_i}$.

\bigskip

From a theoretical standpoint, the construction of $u_i$ from {\bf (Intra$_a$)},
{\bf (Intra$_b$)} and {\bf (Intra$_c$)} is straightforward if the data is consistent.
Indeed, applying {\bf (Intra$_b$)} with $x_i=a^i_j$, $y_i=z_i=\zero_i$ and 
$w_i=\un_i$, together with {\bf (Intra$_c$)} yields
\[ u_i(a^i_j)=k(a_j^i,\zero_i,\un_i,\zero_i)
\]
where $k$ is given by {\bf (Intra$_b$)}. Proceeding in this way ensures uniqueness of
$u_i$.

\medskip

Of course, it is not reasonable to ask a DM to give directly the value of $k$ as
a real number.
The idea of the Macbeth methodology is to ask an information of an ordinal nature 
to the DM. It is well-known from psychological studies that human beings can handle 
at most $5$ plus or minus $2$ items at the same time. The Macbeth methodology
proposes to ask to the DM a satisfaction level belonging to an ordinal scale composed
of $6$ elements:  
\{{\em very small, small, mean, large, very large, extreme}\}$=:\mathcal{E}$. 
If the DM assesses the value of $u_i(a^i_j)$ directly
in this scale, the utility functions will be very rough.

In order to cope with the finiteness of the satisfaction scale $\mathcal{E}$,
we ask the DM to give much more information than just assessing the $u_i(a^i_j)$ 
for $j\in\pa{1,\ldots,p_i}$. In fact, the DM is asked to assess 
(giving a value in the scale $\mathcal{E}$) the difference of satisfaction
$u_i(a^i_j)-u_i(a^i_k)$  between two values $a^i_j$ and $a^i_k$, for any
$j\not= k$ such that $\pp{\zero_{-i},a^i_j}\succ \pp{\zero_{-i},a^i_k}$. The
information asked in practice is thus quite similar to {\bf (Intra$_b$)}.
The advantage of asking $u_i(a^i_j)-u_i(a^i_k)$ is that it leads to a 
redundant information that will enable to compute accurate but non-unique values of $u_i(a^i_l)$
for $l\in\pa{1,\ldots,p_i}$. The second advantage is that it is easier for a human being 
to give some relative information regarding a difference 
(for instance $u_i(a^i_j)-u_i(a^i_k)$) than to give some absolute information
(for instance $u_i(a^i_j)$).

\medskip

There is no unique utility function $u_i$ corresponding to the data composed of 
the values $u_i(a^i_j)-u_i(a^i_k)$ (for $j\not=k$) assessed in the scale $\mathcal{E}$.
All possible solutions are consistent with the given information. In practice, one
utility function is chosen among all possible ones \cite{Bmac}.

The drawback of asking redundant information is that some inconsistencies
may be introduced by the DM. For instance, for $a^i_{j_1}$, $a^i_{j_2}$ and $a^i_{j_3}$ such
that $\pp{\zero_{-i},a^i_{j_1}}\succ\pp{\zero_{-i},a^i_{j_2}}
\succ \pp{\zero_{-i},a^i_{j_3}}$, here is an example of inconsistency~:
the difference of satisfaction degree between $a^i_{j_1}$ and $a^i_{j_2}$ is 
judged {\em very small} by the DM, the one between $a^i_{j_2}$ and $a^i_{j_3}$ is also
judged {\em very small}, and the one between $a^i_{j_1}$ and $a^i_{j_3}$ is 
judged {\em extreme}.  C.A. Bana e Costa, and J.C. Vansnick showed that
inconsistencies are related to cyclones in the preference relation structure \cite{Bmac}.
This property enables to detect and explain all possible inconsistencies \cite{Bmac}.

\subsection{Inter-criteria information}

We could imagine proceeding as for $u_i$ to obtain $\mu$. The DM
would be asked about the difference of satisfaction between the alternatives
$\pp{\un_A,\zero_{-A}}$ and $\pp{\un_B,\zero_{-B}}$ (for $A\not= B$).
Actually, even if this way is possible, it is not generally used because of the
following two reasons. The first one is that it may not be natural for a DM to 
give his preferences on the prototypical acts $\pp{1_A,0_{-A}}$. The second one is
that it enforces the DM to construct a ratio scale over $2^n$ alternatives. 
This requires roughly $4^n$ questions to be asked to the DM. This is too much
in practice.

\bigskip

Conditions {\bf (Inter$_a$)}, {\bf (Inter$_b$)} and {\bf (Inter$_c$)} were 
introduced to show the practicality of the use of the Choquet integral to
model the preference relation $\succ$ on the Cartesian product of the 
attributes.

In practice, {\bf (Inter$_a$)}, {\bf (Inter$_b$)} and {\bf (Inter$_c$)} are
replaced by any classical method designed to deduce the fuzzy measure
from information on acts given in $[0,1]^n$. These acts are directly described by
satisfaction degrees in $[0,1]$ over all criteria. The methods mentioned 
in the introduction are well-suited for performing this step:
we can cite a linear method \cite{Bma2,Bg6},
a quadratic method \cite{Bg4,Bg3} and an heuristic-based method \cite{Bg2}.

\section{Conclusion}

We give in this paper a result that provides some information concerning
the preferences of the DM over each attribute and the aggregation of
criteria, and a priori assumptions leading to the Choquet model. 
The difficulty of the intra-criterion step is that one has to determine the
utility functions (which have a meaning only through the overall model
and thus through the aggregation function) without a precise knowledge 
of the aggregation function (that is determined in the next step).
Let us recall the main property that makes this possible. It is in fact a consequence of
the \axiome{(weak SPL)} property~:
\[ \frac{u(x_i,\zero_{-i})}{u(\un_i,\zero_{-i})}
  = \frac{F_\mu(u_i(x_i),0_{-i})}{F_\mu(1_i,0_{-i})} = u_i(x_i) \ .
\]
Hence, $u_i(x_i)$ is proportional to $u(x_i,\zero_{-i})$ whatever the
value of $\mu$ may be.

\bigskip

Some questions about the
practicality of the information asked to the DM are now raised. 

Note that regarding the questions to be asked to the DM, psychological aspects
must be taken into consideration. For the construction of $u_i$, we propose here
to use the acts $\pp{x_i,\zero_{-i}}$. 
Let us recall that $\zero$ corresponds to the {\em completely
unacceptable} absolute level. As a consequence, since the act $\pp{x_i,\zero_{-i}}$ 
is unacceptable almost everywhere, one could argue that this act can be
considered as almost unacceptable, so that the DM will have
some troubles giving his feeling about the differences between some
$\pp{x_i,\zero_{-i}}$ and $\pp{y_i,\zero_{-i}}$. If this happens, it would
be reasonable to replace the space $X\rfloor_i$ by
\[ X\rfloor_i' := \pa{ \pp{x_i,\un_{-i}} \ , \ x_i\in X_i } \ ,
\]
and thus condition {\bf (Intra$_a$)} by
\begin{description}
\item[(Intra$_a$')] $\forall x_i,y_i\in X_i$, 
$u_i(x_i)\geq u_i(y_i) \Leftrightarrow \pp{x_i,\un_{-i}}\succeq \pp{y_i,\un_{-i}}$.
\end{description}
Then \refe{E5} shall be replaced by
\begin{eqnarray*}
& & \!\!\!\!\!\!\!\!\!\!\!\!\!\!\!\!\!\!
 \frac{F_{\gamma\mu}\pp{\alpha a_i+\beta,(\alpha+\beta)_{-i}}-F_{\gamma\mu}\pp{\alpha b_i+\beta,(\alpha+\beta)_{-i}}}
   {F_{\gamma\mu}\pp{\alpha c_i+\beta,(\alpha+\beta)_{-i}}-F_{\gamma\mu}\pp{\alpha d_i+\beta,(\alpha+\beta)_{-i}}}  \\
 & &  = \frac{a_i-b_i}{c_i-d_i}
\end{eqnarray*}
for all $a_i,b_i,c_i,d_i\in\reel^+$, and for
all $\alpha,\gamma>0$, $\beta\in\reel$.
It can be checked that the rest of the paper remains unchanged, leading also to
theorem \ref{T1}.

If the same phenomenon occurs with $X\rfloor_i'$, the DM satisfies strongly to
the {\em saturation effect}. In this case, it may be more appropriate to model
the DM with the help of a purely ordinal aggregator, like the Sugeno operator.
A construction of a multicriteria decision model in a purely ordinal context,
based on ordinal scales and the Sugeno integral is presented in \cite{Bg7}.

\bigskip

\dep{\large {\bf Acknowledgements}} 

\medskip

The authors are thankful to an anonymous referee for his valuable comments.

\end{document}